\documentclass[manuscript]{acmart}
\AtBeginDocument{%
  \providecommand\BibTeX{{%
    \normalfont B\kern-0.5em{\scshape i\kern-0.25em b}\kern-0.8em\TeX}}}

\copyrightyear{2022}
\acmYear{2022}
\setcopyright{rightsretained}
\acmConference[CHI '22]{CHI Conference on Human Factors in Computing Systems}{April 29-May 5, 2022}{New Orleans, LA, USA}
\acmBooktitle{CHI Conference on Human Factors in Computing Systems (CHI '22), April 29-May 5, 2022, New Orleans, LA, USA}
\acmDOI{10.1145/3491102.3517490}
\acmISBN{978-1-4503-9157-3/22/04}

\newcommand{\red}[1]{\textcolor{black}{#1}}

\begin{document}

\title{``It Feels Like Taking a Gamble'': Exploring Perceptions, Practices, and Challenges of Using Makeup and Cosmetics for People with Visual Impairments}

\renewcommand{\shorttitle}{Perceptions, Practices, and Challenges of Makeup by People with Visual Impairments}

\author{Franklin Mingzhe Li}
\authornote{Equal contribution}
\affiliation{
  \institution{Carnegie Mellon University}
  \city{Pittsburgh}
  \state{Pennsylvania}
  \country{USA}
}
\email{mingzhe2@cs.cmu.edu}

\author{Franchesca Spektor}
\authornotemark[1]
\affiliation{
  \institution{Carnegie Mellon University}
  \city{Pittsburgh}
  \state{Pennsylvania}
  \country{USA}
}
\email{fspektor@andrew.cmu.edu}

\author{Meng Xia}
\authornotemark[1]
\affiliation{ 
  \institution{KAIST}
  \city{Daejeon}
  \country{Republic of Korea}
}
\email{iris.xia@connect.ust.hk}

\author{Mina Huh}
\authornotemark[1]
\affiliation{ 
  \institution{KAIST}
  \city{Daejeon}
  \country{Republic of Korea}
}
\email{minarainbow@kaist.ac.kr}

\author{Peter Cederberg}
\affiliation{
  \institution{Carnegie Mellon University}
  \city{Pittsburgh}
  \state{Pennsylvania}
  \country{USA}
}
\email{pcederbe@andrew.cmu.edu}

\author{Yuqi Gong}
\affiliation{
  \institution{Carnegie Mellon University}
  \city{Pittsburgh}
  \state{Pennsylvania}
  \country{USA}
}
\email{yuqigong@andrew.cmu.edu}

\author{Kristen Shinohara}
\affiliation{
  \institution{Rochester Institute of Technology}
  \city{Rochester}
  \state{New York}
  \country{USA}
}
\email{kristen.shinohara@rit.edu}

\author{Patrick Carrington}
\affiliation{%
    \institution{Carnegie Mellon University}
    \city{Pittsburgh}
    \state{Pennsylvania}
    \country{USA}
 }
 \email{pcarrington@cmu.edu}

\renewcommand{\shortauthors}{Li et al.}

\begin{abstract}
Makeup and cosmetics offer the potential for self-expression and the reshaping of social roles for visually impaired people. However, there exist barriers to conducting a beauty regime because of the reliance on visual information and color variances in makeup. We present a content analysis of 145 YouTube videos to demonstrate visually impaired individuals' unique practices before, during, and after doing makeup. Based on the makeup practices, we then conducted semi-structured interviews with 12 visually impaired people to discuss their perceptions of and challenges with the makeup process in more depth. Overall, through our findings and discussion, we present novel perceptions of makeup from visually impaired individuals (e.g., broader representations of blindness and beauty). The existing challenges provide opportunities for future research to address learning barriers, insufficient feedback, and physical and environmental barriers, making the experience of doing makeup more accessible to people with visual impairments.


\end{abstract}




\maketitle

\section{Introduction}
For the estimated 44\% of the US population that regularly uses cosmetic products \cite{korichi2008women}, a makeup practice constitutes one of the most important avenues toward self-expression and self-care. Despite this fact, there is a near-absent representation of makeup use by people with visual impairments, a worldwide population of at least 2.2 billion \cite{Blindnes59:online}. Makeup practices are equally prevalent and meaningful for people with visual impairments, and embedded in a rich constellation of culture and identity signifiers \cite{Blindnes56:online}. \textit{``When I first lost my eyesight, I was quite sad that I couldn't look in the mirror. Applying makeup is a way that I can control my appearance again,''} said Lucy Edward, CoverGirl's first blind beauty ambassador \cite{Blindnes56:online}. 
Although visually impaired people utilize various tips and tricks in makeup application \cite{BlindGir58:online}, there are profound barriers to participating fully in such efforts of self-expression; purchasing, using, and vetting makeup remains inaccessible \cite{pradhan2021inclusive}. For instance, makeup products have a high reliance on visual information, as many manufacturers do not tactilely differentiate between colors and formulas in their product lines \cite{TheBesta70:online}. Nonetheless, people with visual impairments are attentive to their appearance in the same proportion as sighted peers, especially in spaces that are guided by social norms around makeup \cite{pradhan2021inclusive}. With Edward's testimony in mind, we believe there remain many opportunities for research that furthers the self-expression and personal well-being of people with visual impairments (e.g., \cite{shinohara2011shadow,li2021non}).


To support people with visual impairments in cosmetics, companies like L'Occitane and Proctor \& Gamble have made some progress by attaching braille labels \cite{Beautyis6:online} or non-braille tactile markers \cite{Inclusiv52:online} on their cosmetic packages. \red{Furthermore, existing assistive technologies for people with visual impairments in navigation \cite{ahmetovic2016navcog}, text recognition \cite{deshpande2016real}, object detection \cite{bigham2010vizwiz}, blind photography \cite{brady2013visual,jayant2011supporting}, and color detection \cite{mascetti2016towards} may be helpful for individuals researching cosmetic products online and in-store, as well as using products independently.} Many existing accessible crowdsourcing platforms, such as BeMyEyes \cite{BeMyEyes41:online} and Aira \cite{HomeAira84:online}, connect people with visual impairments to sighted volunteers, which may provide more detailed feedback and instruction on makeup application \cite{HomeAira84:online,BeMyEyes41:online}. 
While research in this area is growing, assistive technologies are still limited in their scope to functional activities---such as browsing online content (e.g., \cite{lazar2012investigating}), or indoor and outdoor navigation (e.g., \cite{ahmetovic2016navcog})---rather than expressive activities.
For instance, even where assistive technologies can be helpful in differentiating basic colors, none are currently sophisticated enough to differentiate tones in a nude eye-shadow palette, much less match to a user's skin color or outfit. Because these existing technology approaches do not support users in all important aspects of the task, we first seek to understand cosmetic practices as an essential mode of expression for people with visual impairments. We then focus on the perceptions and challenges of makeup as an everyday task, which may be enhanced through technological development. As such, the aim of this work is to address the knowledge gap around effectively supporting the expression and independence of people with visual impairments during the makeup process. To achieve this, we explore the following research questions from the perspective of people with visual impairments:

\begin{itemize}
  \item RQ1: What are the existing practices around makeup and cosmetics?
  \item RQ2: What is the importance and perception of doing makeup and cosmetics? And why?
  \item RQ3: What are the existing challenges around makeup and cosmetics? And how could HCI research contribute to solving challenges with makeup and cosmetics for people with visual impairments?
\end{itemize}

To understand RQ1, we first conducted a YouTube video analysis with 145 videos relevant to how people with visual impairments do their makeup. We show nuanced information related to unique practices of makeup by people with visual impairments before, during, and after doing makeup (Section \ref{Results YouTube}). Based on the makeup practices extracted from the video analysis, we then conducted semi-structured interviews with 12 people with visual impairments who have experience with makeup and cosmetics to explore RQ2 and RQ3 in-depth. The interview findings further illuminate individuals' perceptions of makeup; both the meaning of makeup to people with visual impairments and its role in social interactions (Section \ref{results interview}). We then present novel challenges that people with visual impairments are currently facing (i.e., learning barriers, insufficient feedback, physical and environmental barriers) (Section \ref{Results Challenge}). We further discuss the design guidelines and potential opportunities in considerations for assistive makeup technology based on makeup meanings and perceptions, and solutions towards novel challenges (Section \ref{Discussion}).

\section{Related Work}
In this section, we describe related literature to provide a foundation for our work. First, we present existing literature that explores the meaning of makeup and cosmetics to people with visual impairments (Section \ref{Makeup and Cosmetics to People with Visual Impairments}). We then show the enabling and assistive technologies for people with visual impairments that may benefit specific steps of the makeup process (Section \ref{Enabling Technologies for People with Visual Impairments}). In the last section, we describe existing technologies and research for makeup (Section \ref{Existing Technology for Makeup}).

\subsection{\red{The Social Meanings of Makeup for People with Visual Impairments}}
\label{Makeup and Cosmetics to People with Visual Impairments}
Makeup and cosmetics contribute to a consumer culture regarding personal reinvention and transformation, forming rituals that allow people to express different aspects of their identities \cite{TheRiseO94:online}. Makeup may reify gender roles as a rite of passage toward adulthood, or otherwise subvert gender through its applications in queer performance spaces and fashion \cite{jairath1role}. Makeup has also historically functioned as a proxy for professionalism, and upward mobility for female-presenting employees \cite{dellinger1997makeup}---social pressures which have been repeatedly highlighted by critical feminist scholarship \cite{sandra1988foucault}. In response to critiques of socially-mandated makeup use, many makeup-lovers have recently reclaimed their capacity for personal expression. In the widespread 2015 YouTube movement ``The Power of Makeup,'' creator NikkieTutorials challenges ``makeup shaming'' by stating that makeup can instead ``transform you into whom you want to be'' \cite{kennedy2016exploring}. Indeed, individuals may spend significant time developing makeup practices to express evolving standards of beauty \cite{nash2006cosmetics}, or feel themselves as part of a community \cite{gentina2012practice}. Different cosmetic styles could also assist individuals in changing how their facial appearance is externally perceived \cite{liu2016strategy}. Throughout these facets of meaning, it is clear that makeup is a socially complex, embodied practice that reaches far beyond the visual gaze.

For visually impaired individuals, the ways in which makeup furthers identity are equally important. Yet, due to the subjective nature of evaluating one's own visual appearance, conducting a beauty regime may not always be as straightforward as it is for sighted individuals \cite{pradhan2021inclusive}. 
Overall, there is a lack of overall understanding of the social complexity of makeup for people with visual impairments, including its implications for self-perception and in-group status. As such, we explore how individuals' makeup practices change according to common social contexts (e.g., employment, shopping \cite{shinohara2011shadow,li2021choose}), including when people with visual impairments might ask sighted people for assistance with different aspects of the makeup process \cite{pradhan2021inclusive}. We offer this in-depth exploration of makeup-related practices and challenges from the perspective of people with visual impairments.

\subsection{Enabling Technologies for People with Visual Impairments}
\label{Enabling Technologies for People with Visual Impairments}
Existing accessibility issues in makeup and cosmetics influence the ways that blind people navigate their makeup purchases and how they use makeup \cite{pradhan2021inclusive}. In the HCI and Accessibility communities, prior literature explored technologies that assist people with visual impairments in various activities of daily living (e.g., cooking, gaming) \cite{li2021non,guo2016vizlens,carvalho2012audio,kianpisheh2019face,li2017braillesketch}. Existing research such as web accessibility \cite{wang2021revamp,zajicek1998web}, media accessibility \cite{krolak2017accessibility,seo2018understanding}, object recognition \cite{guo2016vizlens,bigham2010vizwiz,chincha2011finding,jabnoun2014object}, tactile design \cite{guo2017facade,he2017tactile}, color identification \cite{neiva2017coloradd}, and indoor navigation \cite{ahmetovic2016navcog,williams2013pray,sato2017navcog3} may be adapted to improve the makeup process for people with visual impairments. We divide this prior work into three categories---online learning by people with visual impairments, visual assistance for interacting with inaccessible interfaces or environments, and physical and tactile design for visually impaired individuals.

\subsubsection{\red{Online Learning by People with Visual Impairments}}
Prior research has explored online accessibility issues and created guidelines for web accessibility. Makeup products are available online both for shopping and browsing, and existing research that focused on making web interfaces accessible \cite{wang2021revamp,zajicek1998web} would assist people with visual impairments in searching for the makeup product that they are interested in. Beyond searching for makeup products, many people with visual impairments now use online media content and MOOCs (i.e., Massive Open Online Courses) for learning (e.g., \cite{krolak2017accessibility,seo2018understanding}). In addition to searching and learning from online media content, existing work on supporting people with visual impairments to share their images or videos may further improve the overall experiences of learning makeup styles and products \cite{bennett2018teens,seo2018understanding}.

\subsubsection{\red{Visual Assistance for Interacting with Inaccessible Interfaces or Environment}}
To interact with inaccessible spaces or interfaces, prior research has used computer vision to identify objects in the environment and components in digital interfaces (e.g., \cite{fusco2014using,bigham2010vizwiz,guo2016vizlens,morris2006clearspeech}). \red{For example, Bigham et al. \cite{bigham2010vizwiz} leveraged crowd workers to help visually impaired individuals recognize various objects. Guo et al. \cite{guo2016vizlens} introduced Vizlens, an accessible mobile app that supports people with visual impairments to interact with inaccessible interfaces through crowdsourcing and computer vision.} Beyond recognizing a specific object or interacting with interfaces, color identification applications (e.g., \cite{neiva2017coloradd}) through computer vision could also be enhanced to assist people with visual impairments in choosing and recognizing products with different colors (e.g., lipsticks). To purchase makeup products, it might be inevitable for people with visual impairments to visit stores (e.g., Sephora) physically. Therefore, prior work on indoor navigation (e.g., \cite{ahmetovic2016navcog,williams2013pray,sato2017navcog3}) could assist people with visual impairments when navigating from a mall entrance to a specific makeup store.

\subsubsection{\red{Tactile Design for Visually Impaired Individuals}}
Beyond using mobile technology and computer vision to support people with visual impairments with daily activities, many researchers explored tactile design for people with visual impairments as a more inclusive approach (e.g., \cite{bliss1970optical,kurze1996tdraw,guo2017facade,he2017tactile}). \red{For example, Guo et al. \cite{guo2017facade} presented a crowdsourced fabrication pipeline to assist people with visual impairments to create physical labels with 3D printed augmentation of tactile buttons. Furthermore, He et al. \cite{he2017tactile} introduced a novel toolchain to create tactile overlays for touchscreens.}\hfill\\

Our research builds upon prior work of \textit{enabling technologies} for people with visual impairments. Although existing technologies might be beneficial toward aiding people with visual impairments with some makeup and beauty steps (e.g., object recognition), there is a lack of domain-specific nuance and capabilities to support visually impaired individuals in makeup.



\subsection{Existing Technology for Makeup}
\label{Existing Technology for Makeup}

Prior research has explored different approaches to reduce the effort of makeup steps through technology, such as providing makeup recommendations \cite{jain2009imaging,nguyen2017smart,ou2016beauty,liu2014wow}, recording and sharing makeup logs \cite{nakagawa2011smart}, improving makeup creativity \cite{treepong2018makeup}, developing instructional makeup videos \cite{truong2021automatic}, enabling interactive makeup experiences \cite{kao2016chromoskin} and supporting marginalized groups in makeup \cite{chong2021exploring}. For example, Jain and Bhatti \cite{jain2009imaging} developed a multimodal cosmetic advisory system that leveraged face recognition and color detection to provide recommendations based on skin tones. To support the creativity of makeup processes, Treepong et al. \cite{treepong2018makeup} introduced an interactive face makeup system that combined 3D face modeling, tangible interfaces, projection mapping techniques, and a drawing system that allowed users to interactively design their makeup that enhanced creativity. As additional examples in learning makeup styles, Truong et al. \cite{truong2021automatic} showed the approach of combining computer vision with transcript text analysis to provide hierarchical tutorials from instructional makeup videos automatically, and Chang et al. \cite{chang2021rubyslippers} used content-based voice navigation for how-to videos in makeup tutorials. Beyond supporting makeup just for the general public, Chong et al. \cite{chong2021exploring} created a makeup recommendation system for transgender individuals through automatic facial recognition systems. Although prior research explored technologies for makeup, these applications and solutions are highly reliant on vision. For example, the interactive face makeup system that Treepong et al. \cite{treepong2018makeup} showed would require certain visual capabilities to use, such as positioning the face in the camera's field of view and placing precise commands in the system. Therefore, there remains work to be done around technologies that benefit people with visual impairments not only in domain-specific ways, but also through accessible means.

\section{YouTube Video Analysis}
To understand the overall makeup practices for people with visual impairments, we choose to first perform a content analysis on YouTube videos. We focus on YouTube both as a rich online space for accessibility research \cite{anthony2013analyzing,li2021non}, and as a hub for a diverse community of makeup users and beauty-focused content creators. Specifically, many blind YouTubers, such as Molly Burke \cite{MollyBur91:online} and Lucy Edwards \cite{LucyEdwa78:online}, create content around the significance of wearing makeup as visually impaired individuals, while also filming tutorials for members of their community. As such, our research in this space allowed us to understand the practices of makeup use directly from an active community of visually impaired creators. \red{Our analysis of YouTube videos on blind makeup provides rich examples of practices that complement and present more diverse and nuanced information that we leverage in our interviews with visually impaired people for in-depth understandings of perceptions and challenges.} In this section, we first show both searching and filtering procedures of the YouTube video analysis (Section \ref{Video Searching and Selection Procedures}). We then describe the coding process and data analysis approach (Section \ref{Data Analysis}). Finally, we present findings from our YouTube video analysis (Section \ref{Results YouTube}). 

\subsection{Method}
\subsubsection{Video Searching and Filtering Procedures}
\label{Video Searching and Selection Procedures}
To broadly understand the overall makeup practices for people with visual impairments, we used the depth-first random sampling method to find videos relevant to the makeup practices of visually impaired people. Similar to prior research of searching steps (e.g., \cite{li2021non,anthony2013analyzing,li2022Exploration}), we combined vision-related keywords (e.g., blind, visually impaired, visual impairment, low vision) and makeup related keywords (e.g., makeup, cosmetics, beauty)) to serve as our searching keywords. To come up with these, researchers first started with basic searches (e.g., blind makeup) then gradually added other combinations of keywords from resulting video titles or descriptions (Table \ref{table:searchterms}). Adopting the approach of Komkaite et al. \cite{komkaite2019underneath}, we terminated the search with the specific keyword if the resulting page no longer included any relevant videos. 

\begin{table}[ht]
\caption{The searching keywords used in the YouTube video analysis. A depth-first random sampling method was used to find videos relevant to makeup practices of visually impaired people.} 
\centering 
\begin{tabular}{|p{8cm}|} 

\hline 
\textbf{Searching Keywords} \\ 
\hline 
Blind Makeup, Blind Cosmetics, Blind Beauty, Blind GRWM, Blind Makeup Tutorial, Blind Beauty Step-by-step, Visually Impaired Makeup, Visually Impaired Beauty, Visually Impaired GRWM, Visually Impaired Makeup Tutorial, Vision impairment Makeup Tutorial, Low Vision Makeup\\
\hline 
\end{tabular}
\label{table:searchterms} 
\end{table}

Our initial video dataset covered 160 relevant videos collected before July 14th, 2021. We then filtered out individual videos if: 1) the video did not include any makeup practice nor tips; 2) both the person applying and the person getting makeup did not have visual impairments (e.g., blindfolded); 3) the video had poor audio and video quality; 4) the video was duplicated; 5) the video was not English-based. After the filtering, we ended up with 145 videos (V1 - V145) in the final dataset. 
Among the 145 videos in our dataset, most videos were uploaded in 2019 (35), while others were uploaded in 2020 (33), 2018 (28), 2021 (15), 2017 (11), 2015 (9), 2014 (7), 2016 (5), 2013 (1), and 2012 (1). \red{The 145 videos were from 77 YouTube channels with the highest number of videos (9) coming from the ``Lucy Edwards'' channel \cite{LucyEdwa11:online}.}  The average length of videos was 904 seconds (ranging from 60 seconds to 4130 seconds).

\subsubsection{Data Analysis}
\label{Data Analysis}

The video analysis mainly consisted of two steps: open-coding \cite{charmaz2006constructing} and affinity diagramming \cite{hartson2012ux}.
To analyze the 145 videos we collected, three researchers first open-coded \cite{charmaz2006constructing} all videos independently. Then the coders met and went through the codes of each video. When there was confusion or conflict on any code, the coder explained that code, and then three coders discussed until they reached a consensus then modified the code. A list of codes was consolidated after the discussion. The same three researchers then performed affinity diagramming \cite{hartson2012ux} to group the codes into candidate themes and refine the themes in terms of definition and naming, iteratively. Finally, we generated four themes and 23 codes. We describe and report the findings based on the four themes in the following subsection.

\subsection{Findings: Practices of Makeup by People with Visual Impairments}
\label{Results YouTube}
\red{In this section, we present the practices of makeup by people with visual impairments from YouTube video analysis. We first present how people learn makeup and available resources. We then describe how visually impaired individuals select and identify makeup products and styles. Afterward, we show the practices often utilized by people with visual impairments in applying makeup. Last, we present how they self-assess their makeup and ask for feedback.}

\subsubsection{\red{Learning Makeup}}
\label{Learning Makeup}

\red{Throughout the makeup learning process, people with visual impairments often refer to online content developed by other blind people. By \textbf{watching makeup videos from content creators with visual impairments}, they were not only motivated (V45, V101), but also learned about existing blindness-specific tips and practices (V21, V45).} For instance, V101 mentioned: \textit{``[I was] obsessed with makeup when I was a young girl and watched YouTube videos about makeup tutorials. That's where the passions come from.''} Makeup videos created by people with visual impairments tended to be more accessible in terms of details in the narration, and most of them explicitly described the product information and the look of the product when it is applied (V34, V35, V51, V66, V92). 

In addition to watching video tutorials from people with similar disabilities, we found that people also prefer \textbf{learning from people with similar complexion or demographic background} to reduce the effort of asking sighted people for feedback and help. In V26, the content creator commented: \textit{``I always like to watch the makeup tutorial videos or product reviews from another content creator who has exact skin color as I do. That reduced my concerns of whether a specific product will look good on my skin color.''}

\subsubsection{\red{Makeup Selection and Identification}}
\label{Makeup Selection and Identification}


\red{We found that visually impaired individuals \textbf{select products with which they can easily discern the amount used} during the makeup process. For example, they used products with pump buttons to measure out how much was coming out (V69, V91). In V69, the content creator commented: \textit{``By using the pump buttons of my CC cream, it allows me to easily track the exact amount I have applied or need to apply.''} Another criterion in choosing products was the \textbf{likelihood of encountering errors due to vision barriers}. In order to prevent errors, people with visual impairments preferred tools with short handles that they can control easily to mitigate limitations in depth perception (V11, V58, V65), old dry mascaras that leave fewer clumps (V5, V66, V71), and sponges that leave fewer streaks on the face (V65, V75). Additionally, V89 further mentioned the importance of having long-lasting solutions to reduce the effort of re-application.}

\begin{figure}[b]
\centering
 \includegraphics[width=1\columnwidth]{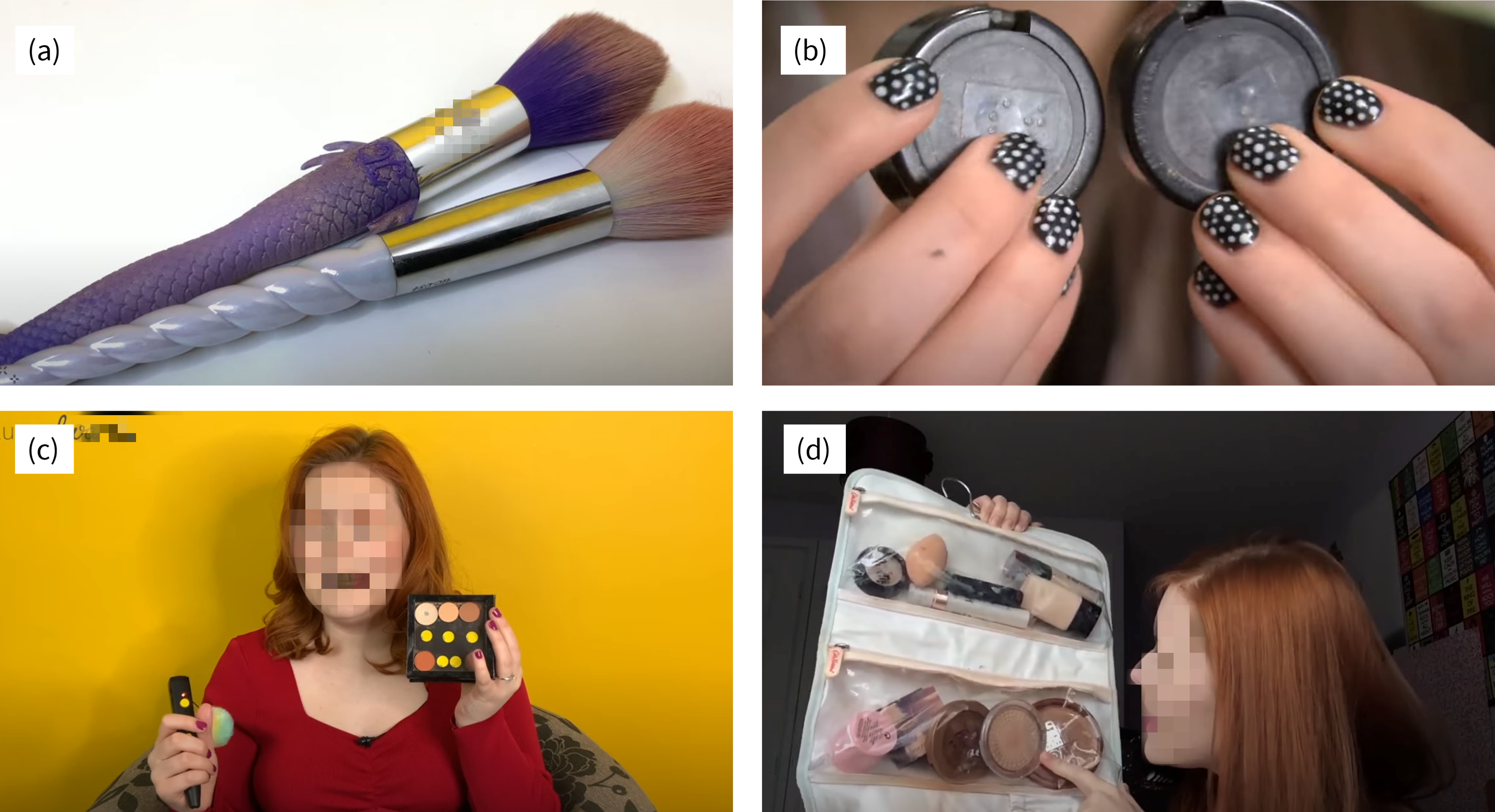}
 \caption{Different ways of distinguishing products. (a) Brushes with different handle shapes. (b) Eyeshadow pans customized with braille labels. (c) Customized audio stickers on an eyeshadow palette. (d) A person shows various products organized in different compartments of a bag.}~\label{fig:distinguish}
 \Description{There are four figures (two rows by two columns) that show different ways of distinguishing products. On the top left, sub-figure (a) shows brushes with different handle shapes. On the top right, sub-figure (b) shows two eyeshadow pans that are customized with braille labels. On the bottom left, sub-figure (c) shows a person holding an audio sticker reader in her right hand and has an eyeshadow palette on her left hand. On the bottom right, the sub-figure (d) shows a person introducing various products organized in different compartments of a bag.}
\end{figure}

We distilled several strategies that people with visual impairments use to distinguish products they own. First, we found that they \textbf{leverage the physical shape of the products to differentiate them} (e.g., V34, V51) (Fig. \ref{fig:distinguish}(a)). The creator of V1 mentioned the importance of the unique tactile shape of the product: \textit{``Any tactile differentiation is a game-changer.''} For products that lack unique physical shape, people with visual impairments added custom tactile markers (Fig. \ref{fig:distinguish}(b)), such as braille stickers (V4, V6, V22, V30, V64), bump dots (V33, V69, V137), or rubber bands (V2). Some people with visual impairments \textbf{use the placement to distinguish makeup products}. They always lined up tools in order (V12, V15, V34, V107, V136) or put them in different compartments of a bag (V26, V39, V40) (Fig. \ref{fig:distinguish}(d)). Additionally, \textbf{other sensory inputs such as sound and smell were utilized to find the right product to compensate the visual impairment}. People with visual impairments added audio stickers (Fig. \ref{fig:distinguish}(c)) to label the product with its name and color (V109), or smelled the product to make sure the correct product was being used (V36, V72). V72 further mentioned: \textit{``I use smell to distinguish makeup products with different scents, especially when they have similar physical packaging, such as lipsticks.''} 

\red{We found people with visual impairments prefer \textbf{asking family members or professionals who know their needs for recommendations that satisfy their personal preferences}. For example, they might consult the makeup store staff in finding the right foundation for sensitive skin and products that are more accessible. V21 commented: \textit{``I always have that one person from the store to help me choose the product that is more accessible or easily identifiable to me, such as products with different shapes or tactile.''}}

\subsubsection{\red{Makeup Application}}

\begin{figure}[b]
\centering
 \includegraphics[width=1\columnwidth]{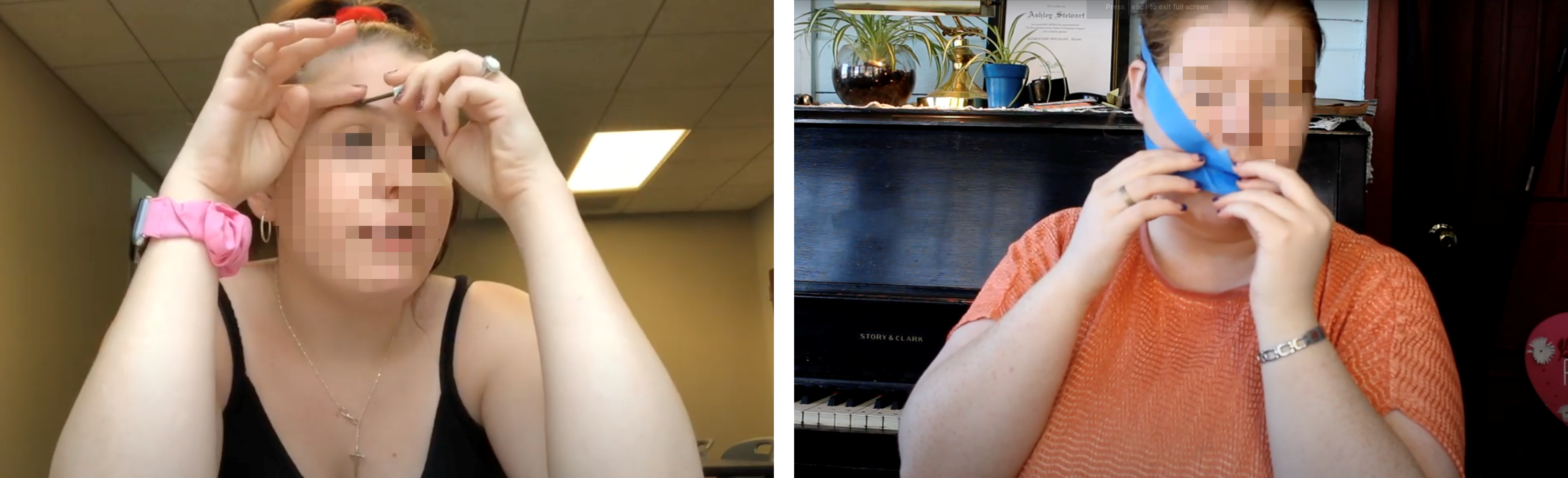}
 \caption{Left: A person uses her hand to guide a brow gel applicator onto their eyebrow. Right: A person uses painter's tape on their cheekbone to help apply contouring.}~\label{fig:application}
 \Description{Two figures. Left: A person uses her hand to guide a brow gel applicator onto their eyebrow. Right: A person uses painter's tape on their cheekbone to help apply contouring.}
\end{figure}

We identified various unique practices from people with visual impairments when applying their makeup. We found visually impaired individuals \textbf{rely heavily on their hands or fingers as an input rather than various makeup tools}. Specifically, we found they usually first apply the makeup product (e.g., eye shadow (V7, V42), foundation (V2, V29), lip products (V38, V96)) on their hand or finger, then apply the makeup on the face. They also \textbf{use their hand or finger to distinguish a specific amount and quantity in makeup}. For example, some people store makeup products in the fridge to distinguish the amount of the product by squeezing them on a hand or finger to feel the temperature difference (e.g., cream (V69)). Furthermore, they often \textbf{used their palm or finger to feel the specific location of the face and guide tools for makeup} (e.g., V92, V99, V100) (Fig. \ref{fig:application}). For instance, V56 showed how visually impaired individuals use fingers to feel where the little hairs are under the eyebrow. Additionally, they often \textbf{clean their fingers or move the products back to the original place after each makeup step to avoid losing track of specific steps and unwanted mess}. 

Besides using a hand or finger as a guide, people also \textbf{used specific tools as guides while applying makeup}. For example, we found that they preferred to have tape- or band-aid guidance, such as using tape to create guides on the face for contouring (V19) (Fig. \ref{fig:application}). Moreover, the reaction from specific tools (e.g., vibration, sound) could provide feedback to visually impaired individuals. One blind content creator commented on her practices of listening to her mascara wand: \textit{``that is how I know that I am grabbing my lashes and it is being coated.''} To ensure more accurate applications in makeup, some visually impaired individuals preferred \textbf{using mini tools or travel-size kits for precision and accuracy} (V15, V31, V61, V116). For example, V61 commented on choosing brushes of small size to reduce the risk of poking eyes.

During the makeup application process, we learned that blind makeup users often rely on memorization to recall their makeup progress. For some specific measurement steps, individuals often \textbf{memorize the number of strokes or how many pumps of makeup to use to control the amount}, such as counting the number of strokes (V19), tapping the brush onto the blush for specific times to pick it up, and memorizing how many pumps of foundation to use. Beyond memorizing the specific amount, we found that they also \textbf{use specific positions on the face as an anchor or reference point during makeup}. For example, people with visual impairments memorize scars with finger-distance from facial parts (V92) and specific places on her face to stipple (V145). Due to the difficulty of memorization, we found that they tend to \textbf{over-do certain steps in makeup}, such as blending (e.g., V8, V89). For example, some blind content creators mentioned that they prefer blending more than sighted people just to make sure it is done (e.g., V4, V30).

In some cases, visually impaired individuals asked other people for help during makeup processes. We found that they sometimes \textbf{ask for help with specific makeup steps}, such as eyeliner or mascara that requires more precision and depth perceptions (e.g., V2, V4, V35). The content creator of V35 commented: \textit{``I normally would have someone to do eyeliner for me because I cannot do it very well myself.''} 

\subsubsection{\red{Self-assessment and Feedback}}
\label{Self-assessment and Feedback}

While applying and finishing their makeup, visually impaired individuals explored various approaches for self-assessing and asking for feedback. First of all, we learned that they often \textbf{use touch and feel to check while doing makeup after specific steps}. For example, V74 mentioned using fingers to feel if the primer is properly applied and blended. For people who can see contours or colors through magnifiers, we found they \textbf{rely on a mirror with magnifier for checking makeup} (e.g., V42, V69). V69 commented: \textit{``I am legally blind and can see color contours with magnifiers, I often use the huge 10x magnification mirrors for help with knowing colors and contours while doing makeup and checking after makeup.''} As cameras are getting higher resolution and magnification, we found that people with visual impairments \textbf{use their phones to take selfies with specialized zooming tools for checking their makeup} (e.g., V41, V56).

After finishing the makeup, we found that people with visual impairments tend to have someone to provide feedback on makeup. First, they \textbf{ask their friends or family members to check their look when done}. In the YouTube videos, they also mentioned that they would prefer not only overall feedback of the makeup, but also the detailed pros and cons about their makeup (e.g., V5, V14). In terms of how they received feedback, they mentioned either using Facetime or asking in person. Beyond asking their friends or family for feedback, very few videos mentioned they would \textbf{use crowd-based apps, such as Be My Eyes or Aira, for feedback} (e.g., V2, V30). However, they mentioned that they still preferred more personalized feedback from their friends and family, if they are available.

\begin{table*}[t]
\small
\caption{Participants' demographic information} 
\centering 

\begin{tabular}{|c|c|c|c|c|} 

\hline 
Participant & Age & Gender & Vision Impairment Description & Learned Doing Makeup \textit{Before or After} Vision Loss\\ [0.5ex] 
\hline 
P1 & 39 & Female & Totally Blind, Acquired (23 yrs) & Before\\
\hline
P2 & 20 & Female & Legally Blind, Acquired (2 yrs) & Before\\
\hline
P3 & 24 & Female & Legally Blind, Congenital & After\\
\hline
P4 & 48 & Female & Totally Blind, Acquired (6 yrs) & Before\\
\hline
P5 & 26 & Female & Legally Blind, Acquired (9 yrs) & Before\\
\hline
P6 & 46 & Female & Totally Blind, Acquired (38 yrs) & After\\
\hline
P7 & 42 & Female & Legally Blind, Congenital & After\\
\hline
P8 & 28 & Female & Totally Blind, Congenital & After\\
\hline
P9 & 22 & Female & Legally Blind, Congenital & After\\
\hline
P10 & 27 & Female & Legally Blind, Acquired (26 yrs) & After\\
\hline
P11 & 33 & Female & Legally Blind, Acquired (3 yrs) & Before\\
\hline
P12 & 43 & Female & Legally Blind, Acquired (10 yrs) & Before \\[0.5ex] 
\hline 
\end{tabular}

\label{table:demographic} 
\end{table*}

\section{Semi-structured Interviews}
\red{Based on the findings describing makeup practices from the YouTube video analysis (e.g., how people with visual impairments learn makeup. identity makeup product, apply makeup, and check makeup quality) (Section \ref{Results YouTube}),} we conducted semi-structured interviews with people with visual impairments to further explore perceptions and challenges of makeup, such as their personal perceptions on the meaning of makeup, and challenges that are related to visual information through various makeup procedures. We first present the methodology of our semi-structured interviews by showing information about participants (Section \ref{participants interview}), study procedures (Section \ref{Study Procedure interview}), and data analysis (Section \ref{data analysis interview}). We then describe our findings for both the perceptions (Section \ref{results interview}) and challenges (Section \ref{Results Challenge}) of makeup shared by people with visual impairments.

\subsection{Method} 
\subsubsection{Participants}
\label{participants interview}
To understand the perceptions and challenges of applying makeup faced by people with visual impairments, we conducted semi-structured interviews with 12 visually impaired volunteers who have experience with makeup and cosmetics. Although our recruitment form asked for participants of any gender identity, all 12 participants who responded identified themselves as female, with an average age of 33, ranging from 20 to 48 years old. Four of them are totally blind, and eight are legally blind. Regarding their first time doing makeup, the distribution is quite even, with half of the participants starting before the loss of vision, and the other half after. Participants were recruited via social platforms (e.g., Reddit, Twitter, Facebook). To participate in our interview, participants had to satisfy the following requirements: 1) be 18 years old or above; 2) have visual impairments; 3) have experience with makeup and cosmetics; 4) be able to communicate in English. The interviews were conducted virtually via Zoom, and each one took around 75 - 90 minutes. Participants were compensated with a \$20 Amazon gift card for completing the interview. The entire recruitment and interview process was approved by the Institutional Review Board (IRB).

\subsubsection{Study Procedure}
\label{Study Procedure interview}
In the interview, we first inquired about participants' demographic information and past experience with cosmetics, such as when and how they first started applying makeup, obstacles they encountered along the way, and how they dealt with them. \red{Our findings in the YouTube video analysis have shown different makeup practices before, during, and after doing makeup (e.g., how to distinguish between products (Section \ref{Makeup Selection and Identification}), and how do they check their makeup (Section \ref{Self-assessment and Feedback})), but these findings did not provide much insight into how people with visual impairments got interested in makeup, and how they figured out what worked best.} We asked participants to discuss their perceptions of wearing makeup both for personal meanings and influence over social interactions (e.g., How does wearing makeup make you feel? What do you think is the role of makeup in society?). Furthermore, we asked participants to share their experiences and any challenges they encounter throughout the entire makeup routine, such as selecting and purchasing products (e.g., Where and how do you purchase makeup products?), learning or following specific practices (e.g., How do you learn makeup styles?), and checking and correcting a makeup look (e.g., How do you check your makeup after it's done?). Finally, we asked participants to describe how they experimented with new makeup, and how makeup could be used to help them express their creativity. 

\subsubsection{Data Analysis}
\label{data analysis interview}
The interviews were recorded and transcribed. Two researchers independently performed open-coding \cite{charmaz2006constructing} on the interview transcripts. Then the coders met together and discussed their codes. When the coders found a conflict, such as a missing code, they explained the rationale for their codes to each other and discussed together to resolve the conflict. Eventually, both coders reached a consensus and consolidated the list of codes. After finishing the coding process, they performed affinity diagramming \cite{hartson2012ux} to group the codes and summarize emerging themes. Overall, we established six themes and 35 codes. The results introduced in the next section are organized based on our six themes.

\subsection{Findings: Perceptions of Makeup by People with Visual Impairments}
\label{results interview}
In this section, we demonstrate the importance, perceived benefits, and concerns of makeup from the perspectives of people with visual impairments. We will first show the perceptions of makeup itself and how that relates to self-expression (Section \ref{Beauty and Aesthetics to Blindness} - \ref{Mutual influences between makeup and confidence}). We then describe the perceptions of doing makeup with respect to social interactions from visually impaired individuals (Section \ref{Social Expectations and Environments to Wear Makeup}). 

\subsubsection{\red{Broader Representations of Blindness and Beauty}}

\label{Beauty and Aesthetics to Blindness}
\red{From the interviews, we learned of misperceptions from the general population about the relationship between beauty and blindness. Over half of our participants reported that they found \textbf{sighted people assume blind people do not ``need'' makeup}, so it is less mentioned or taught at home if none of their family members have visual impairments. According to what P1 and P9 commented, it is not common to see a blind person wearing makeup on social media in general, which creates the stereotype that blind people do not wear makeup. This perspective continues to inform the sparse representation of blind beauty rituals in both media and research, and directly contributes to the lack of accessible tools. P4 further commented on this issue:}

\begin{quote}
    \red{``...Many people that I talked with who are sighted were very surprised that I do makeup. Even my family were not aware of how important doing makeup is in my social life...''}
\end{quote}

\red{We also learned that aesthetic expression is as multifaceted for blind individuals as for sighted peers, and encompasses practices beyond makeup use. For instance, three participants emphasized the importance of \textbf{looking complete and put together} by combining makeup with clothing. However, color and style matching makeup with clothing may also pose additional complexity. P8 commented on this:}

\begin{quote}
    ``...Some people might think doing makeup is already a difficult task for the blind population. However, it is even more complicated because makeup is usually combined with clothes, which added more barriers for us in matching colors and choosing personalized makeup...''
\end{quote}

\red{Despite a lack of representation for blind aesthetic expression, our interviewees asserted that \textbf{blindness does not diminish one's access to beauty}}. P11 further commented:

\begin{quote}
    ``...I have had experiences when planning to use a new product, and others just reacted like it was not necessary to me. Just because we are blind, it does not mean we cannot make ourselves feel beautiful...''
\end{quote}

\subsubsection{\red{Relationship between Makeup and Self-image}}
\label{Mutual influences between makeup and confidence}
\red{We learned that there exists both self-confidence and self-consciousness in doing makeup for visually impaired individuals. In terms of self-confidence, we first realized that there is a positive feedback loop between makeup and confidence for people with visual impairments. Self-confidence is important for visually impaired people living in a world of ableist challenges \cite{shinohara2016self}, five participants emphasized that \textbf{makeup can be a confidence booster} in their daily lives. P1 commented on her feelings of makeup:}

\begin{quote}
    ``...Makeup makes me feel like a better version of myself. And it is not just more confident, it also makes me look more alive and awake...''
\end{quote}

\red{While makeup can be confidence enhancing}, \textbf{having increased confidence can also make someone more likely to play with and enjoy their makeup process} as an act of self-care (P3, P7, P9). Specifically, P7 recounted appearance-based bullying throughout her early years, which prevented her from experimenting with makeup (so as not to further stand out). It was not until P7 was able to regain the confidence to be expressive with her appearance that she began gravitating toward bright makeup. Glitter and blue lipsticks became a source of joy and creativity for her---regardless of others' perspectives. P3 further explained that compliments from other people encouraged her to do more makeup, and she thought of makeup as a way to put her best foot forward because she knows she looks good:

\begin{quote}
    ``...Every time when I hear from other people talking about how beautiful I am with my makeup, I become more excited and confident in trying new makeup styles and willing to spend more time and money on it!...''
\end{quote}

In contrast, six participants also mentioned feeling self-conscious about makeup, which discourages them from doing makeup or makes them become more conservative in trying new things. Three interviewees are \textbf{resistant to being creative in makeup due to concerns of error}. P5 commented on this fact:

\begin{quote}
    ``...I usually just stick to the same color and same techniques I know. I think less is more, and I just want to be comfortable in makeup processes. The consequences of doing new makeup wrong are way more than just following simple makeup routines...''
\end{quote}

We further learned that two participants were often cautious at the beginning and afraid of messing things up. P6 also followed up on the \textbf{discouragement from the complexity of makeup} for people with visual impairments:

\begin{quote}
    ``...There is so much discouragement in makeup. So many steps of makeup are frustrating, especially when I just started to learn makeup. It is also confusing with the tools, the steps and also different makeup products...''
\end{quote}

\red{Finally, we learned that these challenges might be further exacerbated by broader misperceptions of blind makeup use. Our participants expressed that they felt that \textbf{society attaches ableist assumptions to less-than-perfect makeup from an individual with visual impairments}. For example, P8 commented on how imperfect makeup signifies her blindness, rather than a rushed morning:}

\begin{quote}
    ``...When I do my makeup, it needs to be great, because if it is not, it will signify my blindness and give me a lot of pressure. That is why I tried to do things simple and natural...''
\end{quote}

\red{This additional pressure may cause some individuals to remain more conservative in their makeup use, so as to avoid ableist assumptions \cite{campbell2009contours}.}


\subsubsection{\red{Relationship between makeup and social interaction}}
\label{Social Expectations and Environments to Wear Makeup}
\red{In the previous section, we noted how social perceptions influenced how and when participants used makeup (Section \ref{Mutual influences between makeup and confidence}). We also found that participants' makeup played a role in their day-to-day social interactions. In this section, we discuss how visually impaired individuals leverage makeup to control their visibility in sighted settings and engage in the community. We also consider how familiarized surroundings affect one's motivation toward makeup.}

\underline{\textbf{\red{Controlling Visibility:}}} \red{Our participants mentioned makeup could enable them to \textbf{adjust their visibility in social interactions}. \textit{``Makeup helps to control how visible I am!''} said P12. P12 further explained that having a disability often makes her overtly visible to others, such as when wearing sunglasses indoors or using a cane. In contrast to findings mentioned in Section \ref{Mutual influences between makeup and confidence}, makeup also disrupts assumptions about what blind people do. As such, makeup is one especially important way she is able to change her visibility in different contexts:}

\begin{quote}
    ``...Makeup can help to control how I would like to present myself in front of others. For example, if I am happy today and would like to socialize more at a party, I would put on more colorful makeup. Otherwise, I would keep a more natural-looking, and it will be less noticeable...''
\end{quote}

Two participants also commented on \textbf{hesitating to change their makeup or going out without makeup to prevent unwanted attention}. P2 explained: 

\begin{quote}
    ``...I normally keep my appearance the same as before the loss of vision. This makes people focus less on my face and my visual impairments...''
\end{quote}

As a specific example of changing visibility, five participants described how wearing makeup allows them to be \textbf{perceived as professional under different contexts}. P10 emphasized the benefit of makeup in making her feel more professional: 

\begin{quote}
    ``...I have a babyface, and it sometimes causes concerns under some professional contexts, such as interviews. I always wear makeup to feel more professional and confident during formal interviews and presentations...''
\end{quote}

\red{From these participants, we learn that makeup is often one way in which visually impaired individuals may modulate how much attention they receive on account of their blindness. Some individuals choose to shun attention, while others use makeup to harness positive attention, such as during a job interview.}

\underline{\textbf{\red{Community \& Belonging:}}} \red{In our interviews, nine participants mentioned that \textbf{makeup becomes a way to connect visually impaired individuals with other people, and specifically to form friendships with women}. P3 commented:}

\begin{quote}
    ``...Makeup is a big thing for women, I can use makeup to show other people my feelings for each day, and it can become a social medium between other people and me, both sighted and visually impaired populations...''
\end{quote}

P4 further mentioned her experiences of exchanging different makeup products and exploring makeup styles with her friends:

\begin{quote}
    ``...I often exchange my makeup products with my friends, which definitely helps our friendship. We read magazines together and even try new products on each others' faces...''
\end{quote}

P1 talked about how makeup allows her to connect to her sighted female friends and teach them new tips as a makeup enthusiast. P8 further mentioned using makeup to \textbf{guard and enforce her femininity in social interactions}, such as putting on a powerful full-face makeup look during a breast cancer checkup---a moment where her womanhood felt threatened.

\underline{\textbf{\red{Importance of Support \& Motivation:}}} \red{we found that the support and motivation from familiarized surroundings are important to visually impaired makeup users. We found that the \textbf{social pressure of makeup is highly related to the surrounding environments}. For example, P9 is a student in computer science, and she mentioned that the limited number of women in her field made her not that interested or motivated to do makeup every day. P10 also mentioned how pressure from social media made her feel unmotivated:}

\begin{quote}
    ``...The pressure of doing makeup starts at a young age, around middle school. Social media often has perfect makeup, and it becomes an expectation for us. However, this feels like a lot of pressure because of the complexity of makeup for people with visual impairments...''
\end{quote}

In addition to social environments, we found that \textbf{family attitudes also affect how people with visual impairments apply makeup}. Four participants mentioned that their family environments either motivated or discouraged them from doing makeup. While P7's conservative family upbringing caused her to come to love makeup later in life, P11's cousin worked at MAC \cite{MACLipst41:online} and taught her a lot of makeup skills when she was a teenager.

\subsection{Findings: Challenges of Makeup by People with Visual Impairments}
\label{Results Challenge}
\red{We now present key challenges that are currently faced by people with visual impairments, which provide novel opportunities to researchers---learning barriers, insufficient feedback, and physical and environmental barriers.}

\subsubsection{\red{Learning Barriers}}
\label{Learning Barriers}

We previously mentioned that people with visual impairments prefer using online video tutorials to learn new practices and products (Section \ref{Learning Makeup}). \red{However, they also mentioned that they highly rely on friends and professionals for recommendations and feedback due to the limitation of online resources (Section \ref{Makeup Selection and Identification}).} From the interviews, we uncovered various challenges and opportunities to improve video content to better support people with visual impairments in makeup. Most complaints with the video tutorials were about the \textbf{lack of descriptive data in makeup tutorial videos}, and content creators often assumed their audiences did not have visual impairments. For example, P11 commented on the problem:

\begin{quote}
    ``...When I watch YouTube video tutorials, I listen to them and use the actual text descriptions. Because I need more detailed examples, say they are doing a red smoky eye, they are talking about you are going to put on barnyard red and apply to the crease, but they do not say the way how they did it. They assume you can see what you are doing...''
\end{quote}

Moreover, P1 further mentioned the \textbf{overuse of demonstrative pronouns} in videos that confuses visually impaired individuals:

\begin{quote}
    ``...Many YouTubers often use `this' or `that' to refer to a specific product or position. We cannot know what does `apply this thing here' mean?...''
\end{quote}

Beyond watching YouTube videos from blind content creators, our participants also complained about the \textbf{missing detailed information online on how to apply the makeup product step by step.} They proposed that online information should include every detailed instruction that a person might get from actually talking with a makeup specialist from a physical store. P1 explained:

\begin{quote}
    ``...It should include everything that you may imagine you can get from actually talking with a specialist, such as whether it is great for day or night, how long does it last...''
\end{quote}

Other than the needs of descriptions, our interviewees also mentioned the \textbf{inaccessible and irrelevant content from online makeup product pages}, such as moving images (P10) and information overloading from too many ads (P9). Therefore, P11 mentioned that she avoids buying makeup products online, unless she is already familiar with the product. At last, we also learned that \textbf{online reviews or recommendations lack specification and customization}. Six participants mentioned the inconvenience and effort of exploring products and reviews one by one for makeup products and the lack of specifications of reviewer's demographic information, such as skin type and skin tone. P6 further explained her experiences of going over reviews and recommendations:

\begin{quote}
    ``...I like to use online web pages for browsing makeup products these days during the pandemic. However, many of the reviews and recommendations systems do not take my personal information into consideration, such as whether my skin is sensitive? Whether I have a dark skin tone? In addition, many online makeup platforms do not require people to include their skin type or skin tone when they leave a comment on the product review page. This made me frustrated because I do not know whether I should listen to the person or not.''
\end{quote}

Four participants emphasized the \textbf{need for \red{tactile} makeup learning materials for people with visual impairments}. This includes having makeup magazines with tactile versions and instructions on using new tools for people with visual impairments. P10 commented on this:

\begin{quote}
    ``...Many people now still prefer reading beauty magazines on physical covers. However, I barely see any of them having tactile versions available. Again, new instructions on how to apply new tools are still described to sighted populations. It would be nice of makeup companies to include instructions on how blind people may use the makeup tools...''
\end{quote}

\red{Besides the effort of learning and practicing  makeup, we found \textbf{re-learning makeup after the loss of vision is challenging} because people have to abandon their conventional ways of measuring makeup, applying makeup, and performing self-assessment with vision assistance. Also, doing makeup usually requires a long time to practice and adopt a preferred way and style. In our interview, we had six participants who learned makeup before they experienced vision loss.} For example, P11 expressed the difficulties of learning new ways of doing makeup after becoming legally blind and how that is different from before the vision loss:

\begin{quote}
    ``...Applying makeup was a daily routine for me before the loss of vision, I became legally blind three years ago, and it took me a lot of effort from knowing what makeup products are more accessible, which makeup tools are easier for me to use...Of course, it needs a lot of practice and exploration...''
\end{quote}

\red{Moreover, all six participants who learned makeup before the vision loss emphasized the difficulties of maintaining their previous makeup styles right after the vision loss to prevent unwanted attention of the makeup change from the general public.}

\subsubsection{\red{Insufficient Feedback}}
\label{Insufficient Feedback}

\red{From the interview, we further uncovered existing challenges of having feedback---color identification before makeup, tracking specific steps during makeup, and overall feedback after makeup. We found visually impaired individuals not only have a hard time doing color matching before applying makeup, but also \textbf{find it difficult to acquire color information through product descriptions}, especially in online content. For example, both P5 and P10 mentioned that there is a lack of standards for describing colors, \textit{``Nobody understands what the color `moonlight' refers to!''} said P5. P10 further commented on this:}

\begin{quote}
    ``...I wish it could be better for explaining colors. I want more descriptive information. For example, the definition of `neutral tone' might be different between makeup products.''
\end{quote}

From the interview, six participants mentioned the difficulties of identifying color independently. All participants mentioned that they had used color identification apps before. However, four participants complained about the \textbf{technical limitations of recognizing the detailed description of colors} from color identification apps, including exact color code and color shade. P12 commented on this:

\begin{quote}
    ``...I have used different color identification apps. However, none of them can tell me the detailed color information, I understand the light condition might affect the exposure of the phone camera, I would like the color identification app to tell me more information about the color shade or exact RGB value...''
\end{quote}

Beyond lacking a standardized way of describing colors, two participants also mentioned the \textbf{problem of detecting a group of colors} by using color identification apps, such as colors on a palette. P11 further elaborates on this:

\begin{quote}
    ``...Many color identification apps can detect single color from the camera. However, I found it is difficult to recognize colors on my palette. It also needs better ways of describing a series of colors through audio...''
\end{quote}

In addition, eight participants also expressed strong frustration about color blending. We found that people with visual impairments often have \textbf{difficulties in blending several colors and making sure the color blending is even}. For example, P1 and P5 mentioned that it is hard for them to follow a specific order of adding different colors with an exact amount. P9 further explained the difficulty of color blending evenly:

\begin{quote}
    ``...Blending colors on a color pan is already very hard. However, doing makeup often requires me to blend colors on my face. This made me frustrated sometimes because it is very hard to make sure I blended both sides of my face evenly...''
\end{quote}

Besides the challenges with color matching and blending, our participants also mentioned the \textbf{difficulties of following makeup steps or having precise control over tools}, especially for mascara (P4, P5, P11) and eyeliner (P2, P4, P6, P7, P10, P11, P12). P12 complained about the process of applying eyeliners:

\begin{quote}
    ``There are both eyeliner pencils and liquid eyeliners available. However, both ways are extremely hard for me, I am always afraid of having ink in my eyes while using the liquid eyeliner and worried about sticking the eyeliner pencil to my eyes. It is so hard to tell where the eyeliner goes.''
\end{quote}

While applying makeup, our participants also mentioned the \textbf{difficulty of tracking makeup status and memorizing specific steps}, which includes the number of swipes that have been applied on each face (P2, P4, P5) and the layout of makeup products (P5). P5 further commented: 

\begin{quote}
    ``...For blush, I have to exactly count how many times I applied on each face to make sure they are even, I also have to be consistent with the starting and ending point...While using my palette, I have to memorize the layout of all the colors...''
\end{quote}

Due to the difficulties of following and memorizing specific steps, we found some participants avoided doing makeup or following specific makeup steps due to the effort. The most common problem comes from the time consumption of makeup, which includes learning (P1, P5), practicing (P5, P6), and actually applying the makeup in a daily routine (P3). P5 commented on this concern:

\begin{quote}
    ``...Makeup is not just a simple application. It requires you to spend a lot of time to both learn and do your own practices again and again before you can actually wear it to go outside...''
\end{quote}

\red{Our interview findings corroborated our YouTube video analysis about the roles of other people in providing useful feedback about how makeup looked (Section \ref{Makeup Selection and Identification}).} Through the interview with 12 participants with visual impairments, we learned that getting feedback through electronic devices can be problematic when the person they want to ask for feedback is not physically around. First of all, three participants mentioned that \textbf{getting feedback through video calls can be difficult}. P1 explained her difficulties of doing FaceTime with her mom for feedback due to the camera's field of view and light conditions:

\begin{quote}
    ``...When my husband and my mom are not around, I have to call them through FaceTime to get feedback on my makeup. I usually have a hard time zooming in and out for a specific part of my face. And the light condition affects the automatic exposure of my camera which makes my face look a slightly different color...''
\end{quote}

Beyond asking the person they know for feedback, our participants also mentioned that they leverage crowd-based apps, such as Aira \cite{HomeAira84:online} or BeMyEyes \cite{BeMyEyes41:online}, for feedback. However, they mentioned the \textbf{difficulty of finding the right person who is qualified and trustworthy for makeup feedback}. For example, P9 complained that she experienced unknowledgeable volunteers jumping in and providing dishonest feedback to her while using BeMyEyes: 

\begin{quote}
    ``...I have experienced that a random person came without any background knowledge or even vocabulary of makeup while I use BeMyEyes for help. I wish they could pre-filter some people based on my needs...''
\end{quote}

Furthermore, our participants also mentioned that it is difficult to receive immediate feedback from the crowd-based systems. Three participants commented on the \textbf{need for having an automatic makeup feedback system} that can provide specific and personalized feedback to people with visual impairments. P2 further elaborated on this:

\begin{quote}
    ``...I wish there were systems that can alert me if my lipstick goes way out, and can just let me know my makeup looks OK today...''
\end{quote}

\subsubsection{\red{Physical and Environmental Barriers}}
\label{Physical and Environmental Barriers}

\red{In addition to learning barriers and insufficient feedback, there are also physical and environmental barriers, such as the lack of accessible product design, limited technology, and the complexity of different contexts (e.g., makeup shopping aisles).} Seven participants complained about the \textbf{lack of accessible product design}, which includes unreadable product descriptions, lack of tactile features on the package, and uniform shape of different products. P10 commented on the challenges of acquiring product information from physical packaging:

\begin{quote}
    ``...I understand makeup packages are really condensed in terms of the product information. Some products are not even readable through OCR. I wish there could be more approaches, such as having a QR code, to help me get the product information easier...''
\end{quote}

P5 further describes the tactile and shape design of products:

\begin{quote}
    ``...Currently, very few products have tactile dots or markers. We rely on the shape design of makeup products to recognize specific products. However, existing makeup products, such as lipsticks, the same product with different colors all have the same shape and tactile design...''
\end{quote}

We realized that existing technology (e.g., color identifier apps, object recognition apps) is limited for makeup by people with visual impairments, because makeup usually \textbf{requires high precision and adaptability of complex steps}. For example, P3 mentioned that her apps stopped working sometimes, and it is cumbersome to use one technology for a specific makeup step, and additional technologies for other makeup steps. \red{P4 further commented that the limitation of technology forced her to rely on low tech or no tech solutions:}

\begin{quote}
    ``...There are so many limitations of the current technologies. It may just not work under a specific context, or it may break down at any point. This is why I prefer low-tech or no-tech solutions...''
\end{quote}

Moreover, we learned that the \textbf{complexity of different contexts also limits the makeup experiences} to people with visual impairments, especially for shopping experiences. For instance, our participants complained about the inaccessible experiences of exploring beauty products in drug stores due to the lack of in-person support (P7, P12), and the layout of beauty aisles is too complicated for people with visual impairments (P2, P8). P8 further explained:

\begin{quote}
    ``...I always try to avoid in-store shopping for makeup products, especially for stores like Target, I cannot just wander up, and down the beauty aisle, all the products are so close to each other, and sometimes the specific product I want is locked...''
\end{quote}

Finally, participants had \textbf{difficulty modifying their makeup when they were outside}, such as re-applying lipsticks after dinner or adding more eye shadow or mascara (P2, P3). P3 commented on this:

\begin{quote}
    ``...I often have a hard time doing makeup modifications outside of my place. It often requires me to go to the bathroom and ask another lady for help. I wish I had a small magnification mirror with me all the time...''
\end{quote}

\section{Discussion}
\label{Discussion}
\red{From the interview findings, we describe existing perceptions of how people with visual impairments think about the meaning of makeup and how it affects social interactions. We also showed under-explored challenges that are currently faced by visually impaired individuals when doing makeup. Based on these findings, we further discuss the design considerations and potential opportunities for assistive makeup technology based on makeup meanings and perceptions of makeup and the unique challenges faced by people with visual impairments.}

\subsection{\red{Considerations for Assistive Makeup Technology Based on Makeup Meanings and Perceptions}}

\red{Critical disability scholar Alison Kafer argues that there is a ``persistent and pervasive assumption that disabled people's uses of technology are more assistive than creative'' \cite{kafer2019crip}. We have seen this in the relative lack of beauty-related assistive technologies (e.g., problems of detecting a group of colors) (Section \ref{Insufficient Feedback}), which has been further maintained by historically sparse mainstream representation \cite{heiss2011locating}. However, as disability beauty and makeup use is becoming increasingly visible \cite{Disabili8:online}, we see new possibilities for assistive technologies to celebrate and uphold personal expression (Section \ref{Mutual influences between makeup and confidence}). Many assistive technologies are still often evaluated on the basis of their ``ability to move bodies and minds into heightened productivity, efficiency, normalcy, and speed'' \cite{kafer2019crip} rather than how well they attend to the emotional experiences of their users (Section \ref{Mutual influences between makeup and confidence}). Drawing from our interviews, we first recognize that beauty practices can be an emotionally impactful practice for many blind individuals, intersecting ableism, femininity, belonging, and self-image \cite{ettinger2018disability,chandler2018strange}. As such, we see the need for technologies that take into consideration the emotional and social factors of makeup use which may be uniquely exacerbated for blind individuals. Below, we offer a set of questions to help developers build accessible makeup tools toward pleasure, expression, and community support:}


\textbf{Is the experience of using a tool calming and confidence-building?}
We know from our interview data that makeup has a strong and enduring relationship with self-image (Section \ref{Mutual influences between makeup and confidence}). To help users build confidence, consider using product design to offer a calm and luxurious experience and allow users the opportunity to go over content to train up their skills \cite{tran2020paint}. To cut down on overwhelm, automatic suggestions could take the form of capsule makeup kits—neutral starter packs \cite{HowtoBui3:online} with basics that individuals can be assured will work for them in any combination.

\textbf{Does the tool account for social context and personal style?}
We heard from several users about the importance of holistic style, the combination of makeup, clothing, hair, and accessories. This requires that tools be sensitive to a user's context \cite{gulati2021beautifai}, as well as the various elements of their look. Consider crowdsourcing styles that are contextually embedded \cite{MakeupLo86:online}: a smokey eye guide for a night out with friends, a tool that matches lipstick shades to one's accessories, or a trend-inspired clothing and makeup palette.

\textbf{Does the tool make a user feel included in a community or social group?}
\red{It is important to account for makeup as a community practice (Section \ref{Social Expectations and Environments to Wear Makeup}), in addition to individual practice. Because blind individuals rely on friends or influencers of a similar age, skin color, and ethnicity to vet their makeup (Section \ref{Learning Makeup}), a tool could also allow individuals to connect to one another remotely by sharing images and asking for feedback from friends, etc. Consider also how a tool could be used by multiple users, such as a group of friends who are all getting ready together.}

\textbf{Is the tool culturally competent and accounts for unique practices by visually impaired people?}
\red{While there may exist as many application styles as there are users, we have distilled several common practices that may be unique to blind makeup users. Tools can be sensitive to common problem areas such as symmetry, help users count off the number of strokes, or suggest application tips like keeping products cold to help measure the amount used.}

\subsection{\red{Improve Support for Better Makeup Learning Experiences}}

\red{From our findings of both YouTube analysis and interviews, we illuminated that people with visual impairments highly rely on online content for learning purposes, and there exist various learning barriers that are unique to makeup tutorial videos (e.g., overuse of demonstrative pronouns) (Section \ref{Learning Barriers}). We also found that people with visual impairments prefer using online content made by content creators who also have visual impairments. Our research further extends existing research of enabling visually impaired people in content creation, such as photo sharing on social media \cite{bennett2018teens,zhao2017effect}, video editing on YouTube \cite{seo2017exploring,seo2021challenges,seo2020understanding}, and music composing \cite{payne2020blind}. For example, leveraging personal-aesthetic recognition and contextual saliency detection \cite{zhao2017effect} for makeup-related content would reduce the effort of content creators with visual impairments in editing their photos or videos. Moreover, in video editing by content creators with visual impairments, makeup tutorial videos will require more advanced editing \cite{seo2020understanding} to enhance certain products or processes. This would further require more research and development into supporting visually impaired content creators with more specialized editing instructions.}

We also learned that there exist challenges for people with visual impairments to understand video tutorial content for makeup (Section \ref{Learning Barriers}), such as the lack of descriptive data \cite{gagnon2009towards,kobayashi2009providing,natalie2020viscene,pavel2020rescribe,yuksel2020human,liu2021makes} in makeup tutorial videos and overuse of demonstrative pronouns in videos. Specifically for makeup tutorial videos, our participants mentioned that detailed applications and product information are essential for their purposes of watching them (e.g., brand, volume, where and how to apply) (Section \ref{Learning Barriers}), which has different focuses comparing with general videos. For example, a general video may just need audio descriptions such as ``A person is applying mascara.'' However, makeup video tutorials would need detailed descriptions such as ``The person is holding a [brand][product], and apply [product rather than ``this''] from the [distance] to [specific location rather than ``that''] for [number] times.'' Therefore, we suggest future research on designing accessible makeup videos by creating a database \cite{zhao2017effect} to store the detailed information and leverage human-in-the-loop approaches \cite{yuksel2020human} to obtain specific answers to questions from visually impaired individuals as an add-on tool to the video stream synchronously. This database would need to contain not only product-specific keywords, but also domain information relevant to a visually impaired audience, such as describing a mascara wand's shape. \red{Furthermore, the database could also leverage demographic information from content creators (e.g., age, skin type, skin color) \cite{alashkar2017rule} to provide more specialized answers to questions and recommendations to visually impaired makeup users.}


\subsection{\red{Improve Provisions for Makeup Feedback}}


\red{In our study, we showed the existing challenges of insufficient feedback before, during, and after doing makeup (Section \ref{Insufficient Feedback}). According to prior research in blind photography \cite{jayant2011supporting,brady2013visual,gurari2018vizwiz}, future research could explore various "feedback" modes \cite{jayant2011supporting} that are specifically for makeup purposes, such as the number of colors presented in the camera's field of view, light exposure, camera focus, and face area occupied in the photo. This would support people with visual impairments in taking a picture themselves to send to their friends or family members for better color identification and personalized feedback.}

Furthermore, we also learned that individuals with visual impairments have to use BeMyEyes \cite{BeMyEyes41:online} or Aira \cite{HomeAira84:online} for assistance while their family or friends are not around (Section \ref{Insufficient Feedback}). However, our participants mentioned that it is often the case that the volunteers from these services are often not qualified to provide sufficient feedback or that using such services is time-consuming to be practical (Section \ref{Insufficient Feedback}). To reduce the effort of providing feedback, future research should consider creating an automatic feedback system that leverages computer vision \cite{ishikiriyama2017interactive} or other approaches to provide 1) instructional systems that allow people with visual impairments to follow along while their hands are occupied, 2) feedback systems that check the overall quality of the makeup to make sure it would not cause unwanted attention, and 3) feedback systems that provide detailed recommendations on makeup revision and guidelines based on examples provided by the user.

\subsection{\red{Create Workarounds for Physical Barriers}}

\red{From our interview results, we showed that there exist significant barriers to accessing beauty products by people with visual impairments, such as lack of awareness between beauty and disability from the general public, design of uniformly shaped products, inaccessible product descriptions, and complexity of context (Section \ref{Physical and Environmental Barriers}).
To enable easier object detection and information retrieval by people with visual impairments, future product design could leverage RFID \cite{alghamdi2013indoor} or QR code \cite{al2008utilizing,li2019fmt} as assistance to reduce the effort of using OCR to scan the product information on makeup packages (Section \ref{Physical and Environmental Barriers}). Upon having QR codes on product packaging, future research could also leverage augmented reality on mobile devices with audio feedback to assist people with visual impairments in exploring the beauty aisle in stores (e.g., \cite{coughlan2017ar4vi,herskovitz2020making}). Exploring the beauty aisle can be more complex than a clear ground space for augmented reality. This would further bring more consideration of providing additional contextual factors and having more detailed contextual descriptions \cite{herskovitz2020making} in beauty-specific applications.}

\red{Beyond easier recognition and navigation through assistive technologies, there are more opportunities for fabrication research to contribute in building accessible makeup add-ons \cite{KohlKrea38:online} and other tactile design \cite{holloway2018accessible,he2017tactile,hurst2019fabrication}, which further improve the accessibility of doing makeup. For instance, future research could also leverage crowdsourcing and computer vision to provide automated 3D printing to help people with visual impairments to organize and label their makeup products under a reasonable cost \cite{guo2017facade}. Finally, we acknowledge many barriers that consumer-focused technologies alone may not be able to solve. For this reason, we simultaneously encourage product manufacturers to ease the burden on individuals by leveraging differentiated shapes and embossing to create tactile-accessible products.}

\section{Limitations and Future work}
All of the participants we recruited for the interview were either legally blind or totally blind. However, we agree that people with low vision might have different challenges in makeup (e.g., more often using magnifiers or other accessible tools). Furthermore, while all participants who volunteered to our study happened to identify themselves as women, we would be interested in research that accounts for the different preferences, practices, and challenges held by visually impaired participants of other genders. \red{Finally, we only analyzed the practices through YouTube videos that are based in English, but there might exist other makeup practices that were spoken in other languages and happen across different cultures. Therefore, we are also interested in further discovering how cultural differences affect people with visual impairments in makeup.}

\section{Conclusion}
\red{In this paper, we first describe the results of our content analysis with 145 YouTube videos featuring visually impaired individuals doing makeup and illuminate unique makeup practices for people with visual impairments (e.g., over-do certain steps in makeup). We further show the findings from semi-structured interviews with 12 visually impaired participants and highlight the perceptions (e.g., personal meaning and social implications) and existing challenges (e.g., difficulties in blending several colors, precise application, and acquiring feedback) of makeup from their perspectives. We then discuss considerations and potential opportunities for assistive makeup technology based on makeup meanings and perceptions, including supporting better makeup learning experiences, providing makeup feedback to people with visual impairments, and creating workarounds to overcome physical and environmental barriers.} Overall, our findings and discussion shed light on opportunities for future research and development of assistive technologies to empower people with visual impairments in makeup and cosmetics.

\begin{acks}
We would like to thank Yunzhi Li, Frank Elavsky, JiWoong Jang, Michael Xieyang Liu, Zhen Li, Jiannan Li, Yan Chen for their valuable feedback.
\end{acks}
\bibliographystyle{ACM-Reference-Format}
\bibliography{main}


\end{document}